\documentstyle[twocolumn,prl,aps,epsf]{revtex}

\begin{document}
\draft

\twocolumn[\hsize\textwidth\columnwidth\hsize\csname 
@twocolumnfalse\endcsname

\title{Model of surface instabilities induced by stress}

\author{Judith M\"uller and Martin Grant}

\address{
Centre for the Physics of Materials, 
Physics Department, Rutherford Building, McGill University,
3600 rue University, Montr{\'e}al, Qu{\'e}bec H3A 2T8 Canada }

\maketitle

\begin{abstract}
We propose a model based on a Ginzburg-Landau approach to study a strain 
relief mechanism at a free interface of a non-hydrostatically stressed 
solid, commonly observed in thin-film growth.
The evolving instability, known as the Grinfeld instability, is studied
numerically in two  and three dimensions. Inherent in the description is 
the proper treatment of nonlinearities.  We find these nonlinearities
can lead to competitive coarsening of interfacial structures, corresponding
to different wavenumbers, as strain is relieved.  We suggest ways to
experimentally measure this coarsening. 

\end{abstract}


\vskip2pc]

\narrowtext

Elastic effects can strongly influence the morphology of materials, and
hence influence material properties.  If nonequilibrium
elastic energies build up, there are different ways
for solids to release that energy. One is by plastic
deformation, involving dislocations, another is by elastic
deformation, which is commonly seen in thin-film growth.
A non-hydrostatically strained solid which is in contact
with its own melt or vapor can partially release its elastic energy by
a morphological instability at the interface. This strain relief
mechanism gives rise to what appears to be a buckling of the surface into
trenches, or islands, of a particular spacing.  It
was first predicted by Asaro and Tiller\cite{Asaro72}.
Experimentally, it has been observed and studied by Torii and 
Balibar\cite{Torii92} who
strained He$^4$ crystals non-hydrostatically as well by Berr\'ebar
{\it et al.\/}\
\cite{Berrehar92} in polymer crystals. Furthermore, it is
often associated with the dislocation-free Stranski-Krastanov growth
mode (also called island-on-layer mode) of epitaxially grown thin films
as being observed for Ge/Si\cite{Horn77}, InGaAs/GaAs\cite{Guha90} and
InGaAs/InP\cite{Okada97}.  Since the independent rediscovery of the
instability by Grinfeld\cite{Grinfeld93} and Srolovitz\cite{Srolovitz89},
it is often referred to as the Grinfeld instability.  

Several approaches
have been employed to study the instability. They are either based on
static energy minimization calculations by a variational principle
\cite{Grinfeld93}, or on a dynamical interface equation which describes
mass transport, mainly surface diffusion, under the influence of the
chemical potential which comprises surface free energy and elastic
energy
\cite{Srolovitz89,Noziere92,Spencer91,Spencer93a,Spencer93b,Spencer94,Yang93}.  
Linear stability analysis\cite{Srolovitz89,Spencer91,Spencer93a} 
predicts conditions for the onset of instability.
Spencer and Meiron \cite{Spencer94}, 
and Yang and Srolovitz\cite{Yang93} studied the
nonlinear evolution numerically, whereby the surface profile evolved
to smooth flat peaks with sharp deep grooves. These studies
have been limited to dimension $d=2$: Within the interface formulation,
unphysical sharp cusps form within the grooves, leading to numerical stability
problems \cite{Spencer94}.  To make connection to 
experiments, one requires a model which comprises a full nonlinear 
description, and which can be used in three dimensions.

In this paper, we present a Ginzburg-Landau phase-field model of the
phenomena.  An order parameter field $\phi (\vec r)$
determines whether one is in a
hard solid phase, which supports shear, or a soft disordered phase, hereafter
called the liquid phase, which does not.  The position of the interface
coincides with the rapid variation of this field.  Such an
approach has been applied successfully to other
moving-boundary-value problems, such as phase segregation and crystal
growth \cite{Kobayashi93}.  Indeed, our model is numerically
robust, can be implemented in three dimensions, and is readily
generalizable.  We show below that we recover the Grinfeld
instability in linear and highly nonlinear regimes.  We furthermore 
probe the transient dynamics during the morphological instability,
finding that competitive coarsening of interface structures
takes place.  We suggest ways to measure this experimentally.

The physical mechanism for the stress-driven morphological instability can
be understood easily.  A stressed solid can partially relieve its
stress by differentially moving material from valleys to hills,
buckling at a particular wavenumber. In the less 
constrained peaks, lateral relaxation occurs, unlike in the more constrained
valleys. The resulting stress gradient drives the 
instability by creating deeper valleys, thereby increasing the stress 
gradient, and sustaining the growth of the perturbation. At sufficiently 
small length scales, capillarity prevents the formation of sharp cusps.

The model we propose is based on a Ginzburg-Landau approach in which the
elastic strain is a subsidiary tensor variable coupled to a nonconserved
scalar order parameter. This approach is related to that of Onuki
\cite{Onuki89a,Onuki89b}, Onuki and Nishimori\cite{Onuki91}, and 
Sagui, Somoza, and Desai \cite{Sagui94}, which was
used to analyze elastic effects in phase-separating 
alloys \cite{Karim98}. The coarse-grained Ginzburg-Landau free energy is:
\begin{equation}
{\mathcal{F}}(\phi,u_{ij}) = \int_{\vec r} \bigl[f(\phi,u_{ij}) + 
                                      \frac{l^2}{2}|\nabla \phi|^2\bigr],
\label{ftot}
\end{equation}
where integration over 
${\vec r}$ is indicated by the subscript on the integral,
$u_{ij} = \frac{1}{2}({\partial u_i}/{\partial x_j} +
                     {\partial u_j}/{\partial x_i} )$ is the strain and 
$u_i$ is the displacement field. 
The bulk free energy density $f(\phi,u_{ij})$ is given by:
\begin{eqnarray}
f(\phi,u_{ij})& = &\frac{1}{a} \phi^2 (\phi^2 - 1)^2 + 
          \frac{\varepsilon^2}{\kappa} g'(\phi) g(\phi) + \\ \nonumber
        &+& \varepsilon\ g(\phi) \nabla \cdot \vec{u} + f_{el}(\phi,u_{ij}),
\label{fbulk}
\end{eqnarray}
where the first term describes a three-well potential with $\phi = 0$ 
being the liquid and $\phi = \pm 1$ the solid phase, ensuring that the
liquid-solid phase transition is first order.  The potential depths are
determined by the model parameter $a$, which  
together with the parameter $l$ being proportional to the surface tension, 
determines the interfacial thickness. The second term shifts the energy, so 
that, for constant elastic coefficients, solid and liquid are at coexistence.
The convenient choice $g(\phi) = \frac{1}{2}\phi^2 - \frac{1}{4}\phi^4$
guarantees \cite{Kobayashi93} that both bulk phases keep
their equilibrium values $\phi = 0$ (liquid) and $\phi = \pm 1$ (solid).
The coupling constant $\varepsilon$ is related to the externally applied 
stress.  The trace of the strain tensor is
$\nabla \cdot \vec{u}$, and
$f_{el}(\phi,u_{ij})$ is the isotropic elastic free energy\cite{landau86}:
\begin{equation}
f_{el}(\phi,u_{ij}) = \frac{\kappa}{2} (\nabla \cdot \vec{u})^2 + g(\phi)
\mu \sum_{ij} (u_{ij} - \frac{\delta_{ij}}{d} \nabla \cdot \vec{u})^2,
\label{fel}
\end{equation}
where $\kappa$ is the compressibility, and $\mu$ is the shear modulus in the
solid phase alone.  By construction, the shear modulus in the soft
liquid 
phase is zero, whereas it stays nonzero and constant in the hard solid phase.
Since the 
the solid phase supports shear, whereas the liquid phase does 
not, our phase-field order parameter has a transparent meaning in
the context of the liquid-solid transition.

It is reasonable to suppose that the elastic field 
relaxes much faster than $\phi$.  Then the elastic field can be solved
in terms of the order parameter using the  
condition of local mechanical equilibrium:
$\delta {\mathcal{F}}/\delta u_i 
      = \nabla_j \sigma_{ij} = 0$,
where a summation convention over repeated indices is implicit.  The
stress tensor, 
$\sigma_{ij} = 
{\delta {\mathcal{F}}}/{\delta u_{ij}}$, is then given by
\begin{equation}
\sigma_{ij} 
          = (\varepsilon g(\phi) + \kappa \nabla \cdot \vec{u} ) \delta_{ij}
    + 2 \mu g(\phi) (u_{ij} - \frac{\delta_{ij}}{d} \nabla \cdot \vec{u}).
\label{stress}
\end{equation}
The solution of this to first order in the shear modulus is:
\begin{eqnarray}
\nabla \cdot &\vec{u}& = Tr{\bf{A}} - \frac{\varepsilon}{\kappa} g(\vec{r})+\\ 
                                                                   \nonumber
     &+& 2 \mu \frac{\epsilon}{\kappa^2} 
        \int_{\vec{r}\,'} \int_{\vec{r}\,''} 
G(\vec{r},\vec{r}\,') \nabla'_i\ \nabla'_j\ 
        [ g(\vec{r}\,') M_{ij}(\vec{r}\,',\vec{r}\,'') g(\vec{r}\,'') ],
\label{strainTr}
\end{eqnarray}
where $g(\vec r) = g(\phi(\vec r))$, 
\begin{equation}
{\nabla_i u_j} = A_{ij} - ({\epsilon}/{\kappa}) 
  \nabla_i\ \nabla_j \int_{\vec{r}\,'} G(\vec{r},\vec{r}\,') g(\vec{r}\,'),
\label{strain}
\end{equation}
$\nabla^2 G(\vec{r},\vec{r}\,') = \delta(\vec{r} - \vec{r}\,')$, and
$M_{ij}(\vec{r},\vec{r}\,') = \nabla_i\ \nabla_j\ G(\vec{r},\vec{r}\,') -
   ({\delta_{ij}}/{d}) \delta(\vec{r}-\vec{r}\,')$.
In the absence of external strain,
that is $A_{ij} = 0$, the solid will be stressed 
whereas the liquid is stress-free. For a flat surface $\phi = \phi (y)$,
the solution of Eq.\ (\ref{strain})  
in two dimensions is $u_{xx} = u_{xy} = 0$ and
$u_{yy}(y) = - ({\varepsilon}/{\kappa}) g(y)$. Therefore, the
solid will be uniaxially strained with $\varepsilon$ determining the 
strength of that strain.

The elastic field can now be expressed in terms of the order parameter. 
Substituting the solution for the strain field 
gives the free energy in terms of $\phi$ alone.
The long-range character of the elastic field appears through $M$.
Assuming relaxational dynamics, the equation of motion 
is given by:
\begin{equation}
\frac{\partial \phi}{\partial t} = - \Gamma \frac{\delta {\mathcal{F}}}
                                                 {\delta \phi}
     = - \Gamma \Bigl[\frac{f'(\phi)}{a} - l^2 \nabla^2 \phi 
       + \mu \frac{\varepsilon}{\kappa^2} g'(\phi)\ h(\phi)\Bigr],
\label{motion1}
\end{equation}
with $\Gamma$ being the mobility and
\begin{eqnarray}
h(\phi) = &2& \int_{\vec{r}\,'} \int_{\vec{r}\,''} [G(\vec{r},\vec{r}\,')
            \nabla'_i\ \nabla'_j\ 
          M_{ij}(\vec{r}\,',\vec{r}\,'') + \\ \nonumber
 &+& M_{ij}(\vec{r},\vec{r}\,')\ 
          M_{ij}(\vec{r}\,',\vec{r}\,'') ] g(\vec{r}\,')\ g(\vec{r}\,'').
\label{thforce}
\end{eqnarray}

Rescaling length  and time scales  
$\vec{r} \rightarrow {\vec{r}}/{\lambda}$ where $\lambda$ is
a characteristic length scale, such as the wavelength of the perturbation,
$t \rightarrow t{\Gamma}/{\lambda^2}$, 
rescales the parameters to 
$\beta = {\Gamma \lambda^2}/{\Gamma a} $, 
$\epsilon = {l \sqrt{a}}/{\lambda}$ 
and 
$c = \mu a {\varepsilon}/{\kappa^2}$. 
We obtain a dimensionless
equation of motion:
\begin{equation}
\frac{\partial \phi}{\partial t}  
     = - \beta [f'(\phi) - \epsilon^2 \nabla^2 \phi + c\ g'(\phi)\ h(\phi)],
\label{motion2}
\end{equation}
with three parameters $\beta$, $\epsilon$, and $c$, giving the mobility,
capillarity, and shear strength, respectively.

Numerical simulations on a discrete lattice were performed in two and three 
dimensions. Euler's method was used 
for the integration in time. The 
Green function was solved in Fourier space. For all simulations 
presented here the mesh size $\Delta x = 0.01$ or $0.005$, the time step 
$\Delta t = 0.1$ or $0.05$, $\beta = 1.0$, and $\epsilon = 0.01$. 
This choice 
of $\Delta x$ and $\epsilon$ guarantees that the surface is resolved by
at least 8 points.  The parameter set, 
($L_x$, $L_y$, $L_z$, $Y_0$, $c$)
will be specified below, where $Y_0$ gives the initial amplitude of
the surface. Length scales will be measured in units of 
$\Delta x$. Periodic boundary conditions were employed 
in all direction. Thus, the solid was in contact with 
its liquid phase at the bottom and at the top. It was ensured that the solid 
was sufficiently 
thick that the interfaces at the top and bottom acted independently.

A numerical linear stability analysis was performed in two dimensions.
The system was prepared with a small amplitude sinusoidal surface profile 
$Y(x, t=0) = Y_0 sin(q x)$, where $q$ is wavenumber,
and its subsequent evolution was monitored. 
We found that the growth of the amplitude of the Fourier modes 
was initially independent and exponential,
obeying $\exp(\omega(q) t)$, followed by slower constant velocity growth. 
The fitted dispersion $\omega(q)$ is 
consistent with $\omega = A q - B q^2$, where $A \approx 0.2$ and 
$B \approx 25$.  See Fig.~1. 
Perturbations with wavenumber larger
than a critical wavenumber are stabilized by surface tension, whereas 
wavenumbers smaller than the critical wavenumber are unstable, therefore
being a long wavelength instability. The
flat interface however is stable. This agrees with the linear 
stability analysis carried out by Srolovitz\cite{Srolovitz89} for the case 
where evaporation-condensation is the material transport mechanism,
which is appropriate for our model.
\begin{figure}
\narrowtext
\epsfxsize=3in
\centerline{
\epsfbox{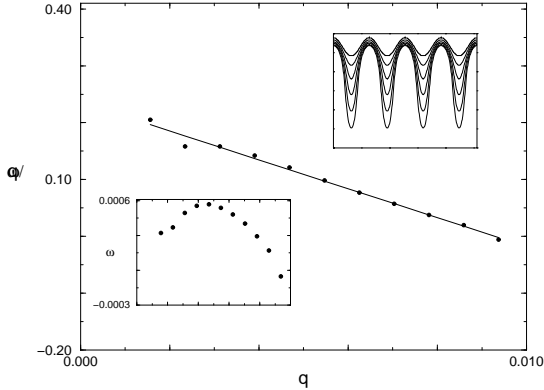}}
\caption{
\noindent
Early-time amplitude of growing mode 
$\omega$ plotted as $\omega/q$ versus wavenumber $q$ to show
that dispersion relation is consistent with   
$\omega=Aq-Bq^2$.  ($L_x=L_y=256$, $c=5.6$,
$Y_0=4$, 1000 time steps).
Lower inset: $\omega$ vs.\ $q$.
Upper inset: 
Time evolution of a configurations in two dimensions sampled every 200
time steps.
($L_x=L_y=512$, $Y_0 = 12.1$, $q = 4/L_x$, $c=5.6$, 
2000 time steps). Linear theory describes only the initial stages of the
instability before asymmetry becomes apparent.
} 
\vskip 0.1in
\label{fig1}
\end{figure}
Linear stability analysis predicts only the condition of onset
of instability. To study the later-stage
morphology, a complete nonlinear description
has to be employed. One advantage of the phase-field description is that 
nonlinearities are taken into account implicitly. A typical set of
configurations is 
shown in Fig.~1. The nonlinear
effect gives rise to
a clear asymmetry between peaks and valleys, wherein deep
grooves appear in the valleys.  This 
behavior has been observed experimentally, as well as in 
previous theoretical studies 
\cite{Noziere92,Spencer93b,Spencer94,Yang93}.
Unlike previous studies, no numerical instabilities limit the study of
the formation of the grooves here.
It is interesting to note that in the early stages of the instability
we can fit the interfacial profile
with a simple function $K = \sum_i a_i(t) Y^i$, where the
curvature $K = Y''(x)/(1+Y'(x)^2)^{3/2}$
is a low-order polynomial function of the height $Y(x)$ of the
interface.

Experimentally, random fluctuations in 
the interface will give rise to the competitive growth of different
structures corresponding to different wavenumbers.
To study this, we prepared the system with an 
interfacial profile consisting of a superposition of $p$ linearly unstable 
modes,  $Y(x) = Y_0 \sum_{i=1}^p \cos(q_i x+\phi_i)$ with 
$q_i < q_c$ and $\phi$ being a uniformly distributed random variable 
in the interval $[0,2\pi]$. 
\begin{figure}
\narrowtext
\epsfxsize=3in
\epsfbox{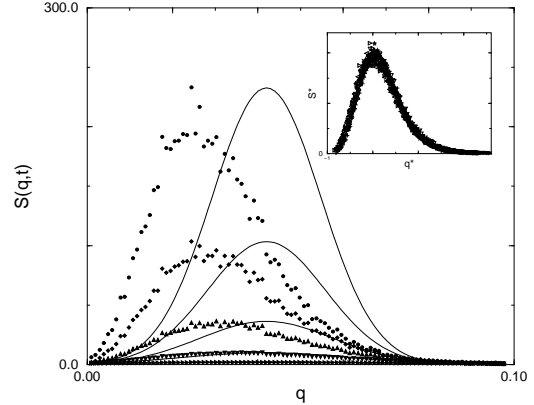}
\caption{
\noindent
Structure factor at equal time intervals.
Solid lines show the structure factor derived from 
a linear Cahn-Hillard-type theory, which only describes the data 
for early times.
For later times, structure factor is consistent with scaling, shown in
inset.
} 
\vskip 0.1in
\label{fig2} 
\end{figure}
We did 100 runs over 500 time steps of a 
two-dimensional system with 100 unstable modes, where 
($L_x$, $L_y$, $L_z$, $Y_0$, $c$)
$=$ (1024, 512, 0, 0.24, 11).
Figure~2 shows the Fourier transform of the equal-time height-height
correlation function, which we shall call the 
structure factor $S(q,t)$.  For early times, 
there is a strong similarity between this behavior and 
early-stage spinodal decomposition in long-range force systems 
\cite{Laradji90};  we show the results of a linear Cahn-Hilliard-type theory 
of the modes in the figure as well. 
Note that the structure factor vanishes for $q \rightarrow 0$ 
due to elasticity, not a conservation law.
For later times, when the linear theory no longer describes the data, 
coarsening is evident: The location of the peak of the structure factor 
$q_{\rm max}(t)$ moves to smaller wavenumbers,
as the peak height increases and sharpens.
The peak height follows $S(q_{\rm max}, t) \sim t^{\alpha}$, where  
$\alpha \approx 2$, while the peak width sharpens with time as 
$w \sim t^{-\gamma}$, where $\gamma \approx 0.5$.
The former dependence is due to the 
total interface length increasing linearly with time for any unstable
wavenumber.  The latter dependence is due to competitive 
ordering between different wavenumbers, analogous to phase
ordering.  Within the accuracy of our study, 
we find that the structure factor shows scale invariance:
$S(q,t)/S(q_{\rm max}, t) = S^* (q^*)$,
where the scaled wave number $q^* = (q-q_{\rm max})/ w$.
See Fig.~2.  Fitting to $S^* \sim (q^*)^\delta$
and $S^* \sim (1/q^*)^\psi$, for small and large $q^*$ respectively,
gives $\delta \sim 1-2$, and $\psi \sim 5 - 6$.

Although these results were obtained in two dimensions, we expect
qualitatively similar results in three dimensions.  To show this, we
simulated a system with $L_x=L_y=L_z=128$, with $z$ being the direction
normal to the surface.
\begin{figure}
\vskip -1cm
 \narrowtext
\epsfxsize=3in
\epsfbox{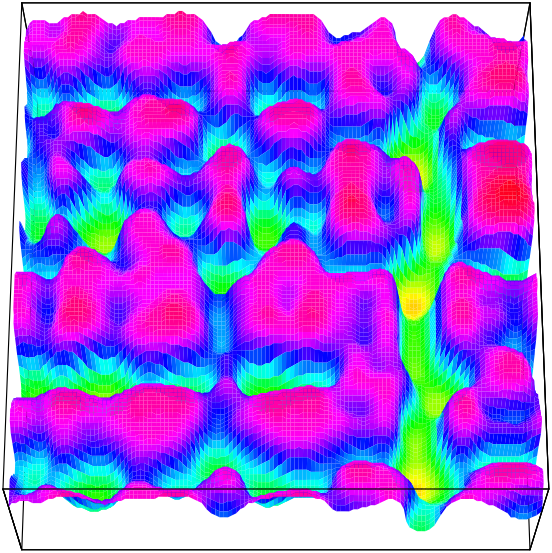}
\caption{
\noindent
Typical configuration in three dimensions after 150 time steps, during
the coarsening regime.
($L_x=L_y=L_z=128$, $Y_0 = 1$, $c=16.7$).
} 
\label{fig3} 
\vskip 0.1in
\end{figure}
Starting with a small amplitude sinusoidal perturbation
in $x$, trenches with sharp deep grooves
form, while a small amplitude sinusoidal perturbation 
in the $x$ and $y$ directions resulted in islands.
The instability is qualitatively the same as in two dimensions.
Starting with a superposition of
unstable modes, coarsening was again observed.  Figure~3 shows the
interfacial profile while coarsening is taking place.
We expect that our results on transient coarsening
phenomena can be observed through microscopy or
by x-ray diffraction \cite{Sinha88}.

To conclude, our model recovers the main features of the Grinfeld
instability.  Our description 
can be easily extended.  Anisotropic effects can included
through the surface tension, the elastic coefficients, or 
the external stress. The effect of
phase separation or of impurities can be studied by coupling an additional
field to the phase field. Instead of evaporation-condensation,
surface diffusion can be chosen as the material transport mechanism,
and, in addition, the influence of a constant flux can be studied.
Finally, we note that in some cases the stress field 
at the groove tip can become so high that
dislocations can be nucleated \cite{Guha90,Dong98}.
To study this, we are presently extending our model by coupling the 
phase field to a dislocation density field.

We thank Karim Aguenaou and Celeste Sagui for useful discussions.
This work was supported by the Natural Sciences and Engineering
Research Council of Canada, and {\it le Fonds pour la Formation de
Chercheurs et l'Aide \`a la Recherche du Qu\'ebec\/}.

\end{document}